\documentclass[graybox, envcountchap]{svmult}

\newcommand{\omegam}{$\Omega_{\rm m}$}
\newcommand{\omegade}{$\Omega_{\rm de}$}

\newcommand{\Hunit}{km/s/Mpc}
\newcommand{\Ho}{$H_0$}

\usepackage{mathptmx}        
\usepackage{amsmath}
\usepackage{amssymb}
\usepackage{color}
\usepackage{helvet}          
\usepackage{courier}         
\usepackage{dirtree}

\usepackage{makeidx}        
\usepackage{graphicx}        
\usepackage{subfig}

\usepackage{multicol}        
\usepackage[bottom]{footmisc}

\usepackage{hyperref}        
\hypersetup{colorlinks=true,urlcolor=blue}

\usepackage[misc]{ifsym}

\makeindex             

\begin{document}


\title{Addressing the Hubble tension with cosmic chronometers}
\author{Michele Moresco}
\institute{Michele Moresco (\Letter) \at Dipartimento di Fisica e Astronomia ``Augusto Righi'', Universit\`a di Bologna, Viale Berti Pichat 6/2, I-40127, Bologna, Italy, \email{michele.moresco@unibo.it}
}
%
%
\maketitle

\abstract{Twenty years after the discovery that the expansion of the Universe is accelerating, a new finding is now challenging our understanding of the cosmos. Recent studies have shown that the Hubble constant, the speed of expansion measured today, provides values in significant tension when measured from the Cosmic Microwave Background in the primordial Universe or from Cepheids and Supernovae Type Ia in the local Universe. Whether this tension is hinting towards new physics or some issue in the measurements, is still under debate; but it is clearly calling for new independent cosmological probes to provide additional pieces of evidence to solve this puzzle.
This chapter introduces the method of cosmic chronometers, a new emerging cosmological probe that can provide cosmology-independent estimates of the Universe's expansion history. This method is based on the fact that the expansion rate of the Universe can be directly derived from measuring how much the Universe has changed in age between two different redshifts, i.e. by estimating the slope of the age--redshift relation.
First, the main ingredients of the method will be discussed, presenting the main equations involved and how to estimate from the observables the needed quantities. After, it will be presented how to reliably select a sample of tracers to map the age evolution of the Universe coherently. Next, different methods to robustly measure the differential age of a population, the fundamental quantity involved in the method, will be reviewed. Finally, the main measurements obtained will be presented, providing forecasts for future surveys and discussing how these data can provide useful feedback to address the Hubble tension.}

\section{Introduction}
\label{sec:intro}
The discovery of the accelerated expansion of the Universe \cite{SupernovaSearchTeam:1998fmf, SupernovaCosmologyProject:1998vns} has been a major revolution in modern cosmology, introducing a shift in the paradigm of the Universe considered until the end of the XX century. This discovery required scientists to introduce the presence of an unknown form of energy driving this acceleration (the ``dark energy''), or to postulate that the laws of General Relativity as known would break at some large cosmological scales.

In this context, finding diverse independent ways to measure the expansion history of the Universe became one of the main goals in astrophysics, since it could provide fundamental inside into understanding the nature of dark energy, or if modified gravity models are a better explanation. The quantity that measures how the Universe is expanding as a function of time (or equivalently redshift) is the so-called Hubble parameter, $H(z)$. It is typically useful to link this quantity to the scale factor $a(t)$ (providing the relation between physical and comoving distances $r_{\it phys}=a(t)r_{\it com}$) with the equation:
\begin{equation}
 H(z)=\frac{\dot{a}}{a} \; .   
 \label{eq:Hz}
\end{equation}

Several cosmological probes have been introduced during these years to address these open questions, measuring how the Universe is expanding and modeling it through cosmological parameters to characterize the behavior of dark energy and dark matter. Due to methodological, technical, and theoretical advances, several probes have reached nowadays a maturity that allowed them to be considered ``standard'' probes, among which e.g. the Cosmic Microwave Background (CMB), Type Ia Supernovae (SNeIa), Weak Lensing (WL), and Baryon Acoustic Oscillations (BAO) (for a detailed review, see, e.g. \cite{Huterer:2017buf}). Cosmological missions have been devised and carried out constantly improving the accuracy and precision of the measurements, pushing all these probes almost to their limits (e.g. Planck \cite{Planck:2018vyg} for the CMB, Pantheon+ \cite{Brout:2022vxf} for SNeIa, and BOSS and eBOSS \cite{Ross:2020lqz} for BAO). It is worth mentioning among these the just launched ESA mission {\it Euclid} \cite{EUCLID:2011zbd}, which will provide the most advanced results for BAO and WL.

Recently, significant attention has been given also to the determination of the local value of the expansion rate, the Hubble constant ($H_0=H(z=0)$), that since first Hubble's measurement \cite{Hubble:1929ig} in 1929 has improved its accuracy by orders of magnitude. In particular, with the observational and theoretical of recent years both direct measurements from the local Universe considering Cepheids and SNeIa (SH0ES, \cite{Riess:2021jrx}) and indirect measurements from the Early Universe based on the analysis of CMB (Planck, \cite{Planck:2018vyg}) have reached the golden goal to constrain \Ho~ with an uncertainty of 1 \Hunit~ and less. However, the results obtained from the two probes presented a tension increasing with the decreasing errors of the two analyses, reaching now a 4-5$\sigma$ disagreement called the ``Hubble tension'' (for a comprehensive review, see \cite{Verde:2019ivm,Abdalla:2022yfr}).
While the origin of this tension is still under investigation, as done in the past it is fundamental now to go beyond the standard probes, which are reaching their intrinsic limits, and explore new solutions to constrain \Ho, providing new and independent measurements that could help to shed light in this debate between early- and late-Universe analyses.

A typical pattern of the various cosmological probes previously discussed is the search for a standard property of objects (luminosity, length) that allows us to decouple in the observable the effect due to the intrinsic evolution and to the expansion of the Universe, making it possible to measure the latter. Historically, standard candles (SNeIa) and standard rulers (BAO) have been amongst the most successful ones, but in this context it is possible to exploit also the time. Using the oldest objects in the Universe to set constraints on its age and on cosmological parameters is not new in the literature (e.g., see \cite{Jimenez:1996at, Krauss:2003em, Jimenez:2019onw, Cimatti:2023gil}, and for a review \cite{Moresco:2022phi}). Moreover, given how the age of the Universe scales as a function of redshift:
\begin{equation}
\mathrm{t(z) = \int_0^z \frac{dz'}{H(z')(1+z')}}.
\label{eq:agez}
\end{equation}
finding the oldest objects at each redshift it is possible to use them to constrain the Hubble parameter inside the integral; for some applications of this method, see, e.g., Refs.~\cite{Vagnozzi:2021tjv, Borghi:2021rft, Tomasetti:2023kek}.
In this way, we would be searching for standard clocks, objects whose absolute ages can be used to constrain cosmological models. This approach has, however, some downsides. First of all, robustly determining absolute ages is quite difficult, requiring an absolute calibration that increases the total error and being prone to not negligible systematic effects. Moreover, as can be evinced from Eq.~\ref{eq:agez}, with this method we measure an integral of $H(z)$\footnote{The same holds for SNeIa, that measure the luminosity distance $d_L(z)$.}, and typically one has to assume a cosmological model, substitute it in Eq.~\ref{eq:agez}, and constrain its parameters. The method is, therefore, not purely cosmology-independent, since we have to assume each time a particular cosmological model.

With the cosmic chronometers method, we can improve on this approach, bypassing all these issues.

\section{The cosmic chronometers method}
\label{sec:method}

The idea behind the cosmic chronometer method has been introduced by Ref.~\cite{Jimenez:2001gg}. Under minimal assumptions, namely just a Friedmann-Lema\^{i}tre-Robertson-Walker (FLRW) metric, given the relationship between the scale factor and redshift $a(t)=1/(1+z)$, the Hubble parameter can be derived directly from Eq.~\ref{eq:Hz} as:
\begin{svgraybox}
\begin{equation}
H(z) = -\frac{1}{1+z}\frac{dz}{dt}
\label{eq:CC}
\end{equation}
\end{svgraybox}
This means that by measuring the differential age of the Universe (how much the Universe has aged between two redshifts) it is possible to obtain a direct and model-independent determination of the expansion rate $H(z)$. The main difference here is that instead of looking for some standard clocks, we will be looking for {\it standard chronometers}, a homogeneous population of objects with a synchronized star formation, i.e. whose clocks started ``ticking'' at the same time and that are therefore optimal tracers of the differential age evolution of the Universe. There are many advantages in this technique with respect to using absolute ages: {\it (i)} the measurement of differential ages has been proven to be more accurate and robust than absolute ages, since many systematic effects in age determination are minimized when estimating the differences $dt$; {\it (ii)} the differential approach allows us to minimize many other systematic effects that might affect this type of analysis (e.g. progenitor bias, rejuvenation, ...) since for the application of the method we are only interested in the homogeneity between the two redshift bins where the quantity $dt$ is estimated; and {\it (iii)} it provides a direct determination of the Hubble parameter without relying on any cosmological assumptions (apart the cosmological principal and a metric). A detailed review of cosmic chronometers is provided in Ref.~\cite{Moresco:2022phi}.

\begin{svgraybox}
The pillars of the cosmic chronometers method are the following:
\begin{itemize}
\item {\bf selection of a population of optimal cosmic chronometers.} As can be evinced from Eq.~\ref{eq:CC}, the first point to address is finding a population of objects able to trace homogeneously how much the Universe has aged between two redshifts. We also need to ensure a way to maximize the homogeneity and purity of the sample, devising criteria to minimize the potential contamination from outliers. We describe the selection process in Sect.~\ref{sec:sel}.
\item {\bf robust measurement of the differential age $\mathbf{dt}$.} While in Eq.~\ref{eq:CC} there are formally two observables ($dz$ and $dt$) the advances in spectroscopic surveys makes the measurement of $dz$ remarkably accurate when a spectroscopic redshift is available (typically $\delta z/(1+z)\lesssim10^{-3}$, see e.g. \cite{Moresco:2012jh}). Therefore, for a proper application of the method, we need to find ways to obtain a robust and unbiased determination of $dt$, as we will describe in Sect.~\ref{sec:age}.
\item {\bf assessment of the statistical and systematic effects.} For a proper application of the method, we need to consider carefully and assess the various sources of errors -- statistical and systematic -- and include those in the total error budget with a proper covariance matrix. We will discuss this at the end of Sect.~\ref{sec:age}.
\end{itemize}
\end{svgraybox}

\section{Selecting an optimal sample cosmic chronometers}
\label{sec:sel}

The most elementary objects in the Universe that we can date and that can be found from the local Universe up to high redshifts are, of course, galaxies. To apply the cosmic chronometer method, the idea is therefore to find at each redshift the oldest population of galaxies available. These will, of course, have an offset with respect to the age of the Universe given the fact that their redshift of formation is not infinite, but as long as our selection is homogeneous this will not impact the analysis, since the differential age $dt$ will not be affected by this issue.

The first and more straightforward solution that can be explored is to avoid any selection, collect at each redshift all available galaxies, and determine the upper envelope of the distribution, selecting the eldest crust of objects. This approach (applied, e.g., in \cite{Jimenez:2001gg, Jimenez:2003iv, Simon:2004tf, Moresco:2012jh, Jimenez:2023flo}) while having the advantage of relying on the easiest possible selection, has the problem that to be applied we need a complete sample with a very high statistic (typically $O(10^3)$ galaxies, see \cite{Jimenez:2023flo}) to robustly detect the upper envelope.

A viable and more feasible alternative suggested in the literature is instead to select the most massive ($log(M/M_\odot)$>10.5-11) and passively evolving galaxies. There is a large literature confirming that these objects have formed at extremely high redshifts ($z>2-3$, \cite{Daddi:2005uv, Choi:2014vfa, McDermid:2015ska, Carnall:2018, Estrada-Carpenter:2019, Carnall:2019}) over a very short time-scale ($\tau<0.3$ Gyr, \cite{Thomas:2009wg, McDermid:2015ska, Citro:2016, Carnall:2018, Jiao:2022aep, Tomasetti:2023kek}). These galaxies have been found up to $z\sim2.5$ \cite{Daddi:2003hu, Fontana:2006xg, Caputi:2012uf, Muzzin:2013yxa, Merlin:2019}, and with the advent of JWST this boundary has been pushed up to $z\sim5$ \cite{Carnall:2023}. These systems, having formed most of their mass at very high redshifts, in a very quick episode of star formation, and having mostly exhausted their gas reservoir are expected to evolve passively as a function of cosmic time; therefore, they represent the best cosmic chronometers. This is confirmed by the fact that their stellar metallicity is found to be very stable as a function of redshift, around solar to slightly over-solar values at $0\lesssim z\lesssim2$ \cite{Gallazzi:2005df, Onodera:2012en, Gallazzi:2014rca, Conroy:2013iaa, Onodera:2014rpa, McDermid:2015ska, Estrada-Carpenter:2019, Kriek:2019}. In addition, it has also been found that more massive galaxies undergo a faster process of formation with respect to less massive ones, forming also at earlier cosmic times. This scenario, referred to as mass downsizing, suggests that most massive passive galaxies are therefore ideal cosmic chronometers, being at each redshift the oldest objects and having a much more synchronized time of formation \cite{Heavens:2004sr, Cimatti:2004gq, Thomas:2009wg}.

\begin{figure}[t!]
\includegraphics[width=\textwidth]{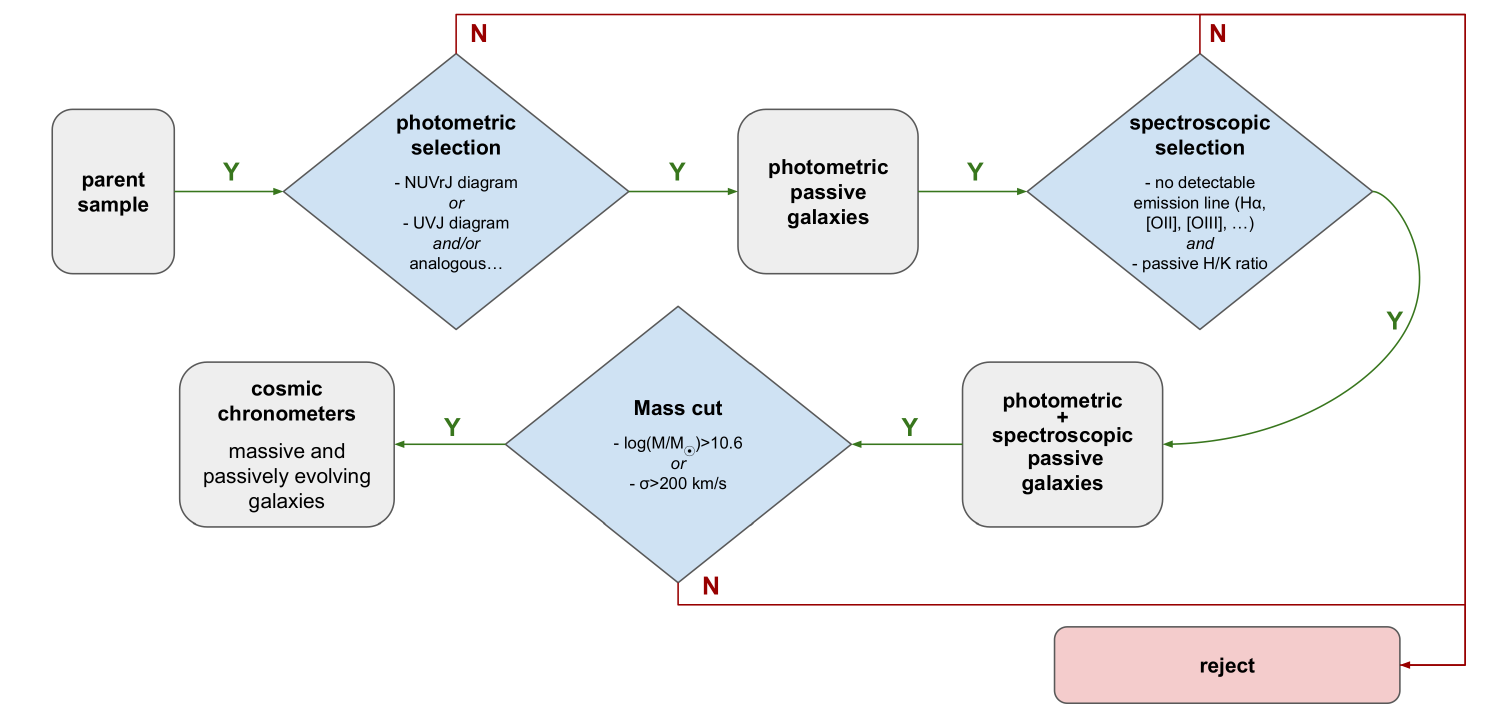}
\includegraphics[width=0.47\textwidth]{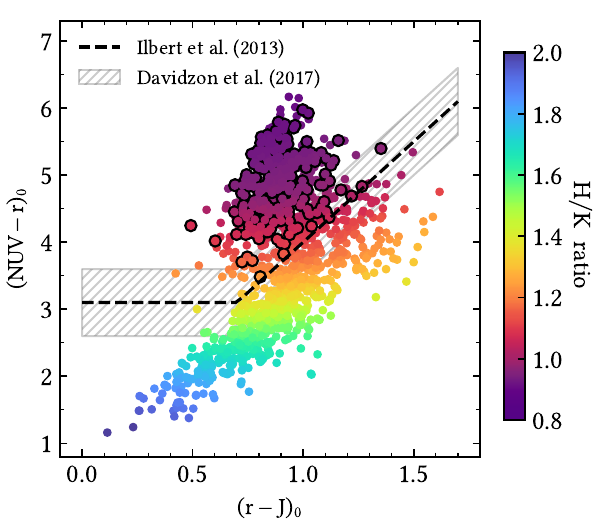}
\includegraphics[width=0.53\textwidth]{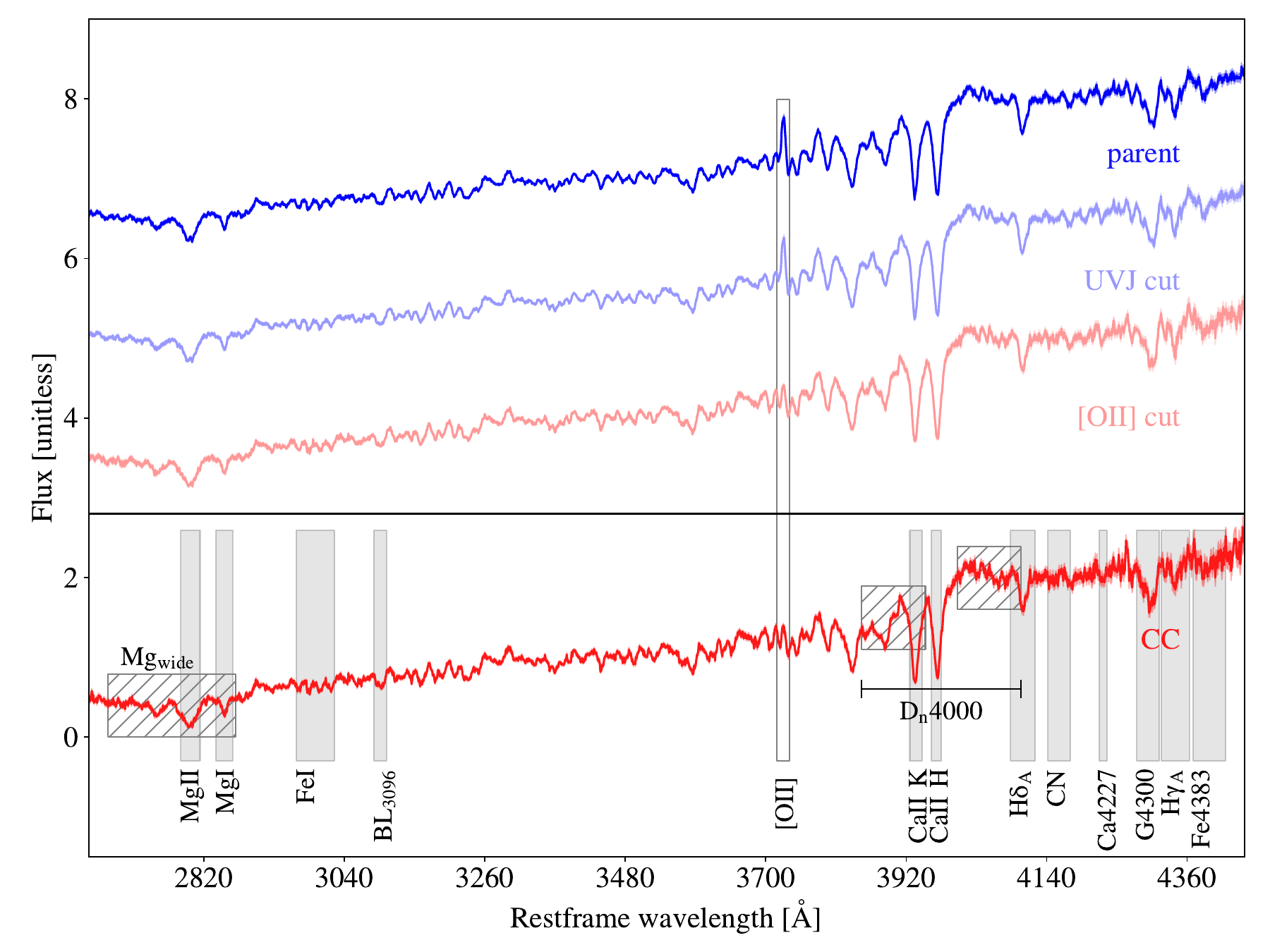}
\caption{Cosmic chronometer selection process. In the upper panel is shown the optimal workflow to select a pure sample of cosmic chronometers (reproduced from \cite{Moresco:2022phi}). The lower left plot show a NUV-r-J color-color diagram, colored as a function of the CaII H/K ratio The dashed line displays the theoretical lines separating passive galaxies (upper part) from star-forming galaxies (lower part), showing the very good correlation with the CaII H/K ratio, demonstrating how it is an excellent diagnostic to select passive galaxies. The right panel shows the effect of different selection criteria on the spectra of passive galaxies. The sample has been obtained from Ref.~\cite{Tomasetti:2023kek}, showing, from top to bottom, the stacked spectra of passive galaxies selecting only on photometry (parent), adding a U-V-J color-color cut ($\rm UVJ_{cut}$), adding a spectroscopic cut on [OII] emission lines ($\rm [OII]_{cut}$), and finally a cut on CaII H/K ratio (CC). It is evident how combining more and more restrictive selection criteria, the emission lines progressively disappear from the stacked spectra, displaying at the end the characteristic features of a purely passive population.}
\label{fig:sel}
\end{figure}

Different definitions have been proposed in the literature to select passive galaxies, based on photometry, spectroscopy, physical properties, and/or morphology. However, it has been also proven that a single selection criterion is not sufficient to obtain a sample with high purity and that depending on the adopted criterion the contamination by young/star-forming outlier can be as high as 30-50\% \cite{Moresco:2013wsa} (see, e.g., the lower-right panel of Fig.~\ref{fig:sel}). This effect can be significantly mitigated by combining different selection criteria, following the workflow proposed by Ref.~\cite{Moresco:2018xdr} and shown in the upper panel of Fig.~\ref{fig:sel}.

In summary, for an optimal selection of cosmic chronometers, it has been demonstrated that the following multiple selection criteria should be combined.
\begin{itemize}
\item A {\it \bf photometric selection}. It has been shown that the wider the photometric coverage, the better passive and star-forming galaxies can be separated. From this point of view, the near-UV and bluer optical bands are crucial, in particular to detect possible contamination by a younger population (0.1-1 Gyr). So far, the NUV-r-J color-color diagram \cite{Ilbert:2013bf} has been identified as the criterion that allows the best identification of passive galaxies (being minimally sensitive to contamination due to dust), but valid alternatives are also the U-V-J diagram \cite{Williams:2008zd} and the NUV-r-K \cite{Arnouts:2013kba}; as another option, it is also possible to assess the galaxy type modeling the full spectral energy distribution (as done e.g. in \cite{Ilbert:2008hz,Zucca:2009nr}).
\item A {\it \bf spectroscopic selection}. Different spectroscopic features have been identified as ideal tracers of ongoing star formation, or of potential contamination by a younger population; among those, the most used are the [OII]$\lambda$3727, H$\beta$ ($\lambda=4861$~\AA), [OIII]$\lambda$5007, and H$\alpha$ ($\lambda=6563$~\AA) emission lines.  Different approaches can be used to reject objects with significant emission, either adopting a cut on the equivalent width of the line (a typical threshold is EW$<$5\AA \cite{Mignoli:2008yz, Moresco:2012jh, Borghi:2021zsr}) or on the signal to noise ratio \cite{Moresco:2016mzx,Wang:2018}). These cuts, while increasing the purity of the sample, are not optimized to detect the presence of potential sub-dominant contamination by a young population, which might affect the oldest galaxies e.g. via episodes of mergers. 
Recently, Ref.~\cite{Moresco:2018xdr} proposed a novel spectroscopic indicator able to precisely assess this level of contamination. The ratio between the CaII H ($\lambda=3969$~\AA) \& K ($\lambda3934$~\AA) lines was found to be maximally sensitive to this ratio, since the presence of a younger population, characterized by a significant H$\epsilon$ ($\lambda=3970$~\AA) absorption lines, affects the CaII H/K inverting its typical ratio. For this reason, it is a powerful diagnostic for the presence of contamination, and it has been shown to correlate extremely well with other indicators such as star formation rate, NUV, and optical colors \cite{Borghi:2021zsr}. As an example, in the bottom-left panel of Fig.~\ref{fig:sel} it is shown how the CaII H/K ratio perfectly map the region of passive galaxies.
Depending on the data availability, it is also advisable to combine several of these indicators to maximize the purity of the sample. In the end, in the spectra of the galaxies ideally should not be visible any residual emission lines, as shown by the bottom-right plot of Fig.~\ref{fig:sel}.
\item A {\it \bf cut in stellar mass or stellar velocity dispersion}. As previously discussed, different studies have shown that the most massive galaxies are the ones that formed before and over the shortest timescale\cite{Heavens:2004sr, Cimatti:2004gq, Thomas:2009wg}; a cut in stellar mass (typically $\log(M/M_{\odot})>$10.6--11) ensures the synchronicity of the sample selected, and, if not available, it can equivalently be done on the stellar velocity dispersion of the galaxy, since it presents a good correlation with the stellar mass.
\end{itemize}
The combination of these criteria has been demonstrated to be effective to maximize the purity of the sample, minimizing possible residual contamination of a young and/or star-forming subdominant population. However, it might not be always feasible to combine multiple criteria, due to data quality or availability. In that case, it is crucial to try to characterize the purity of the sample and properly take into account potential issues in the total covariance matrix as discussed in Sect.~\ref{sec:age}.

\section{Measuring the differential ages}
\label{sec:age}

Measuring the age of a stellar population is not a trivial task. While encoded in the integrated spectrum of the galaxy analyzed, this information presents some degeneracy with other physical properties, the most known being the age-metallicity degeneracy \cite{Worthey:1994iw, Ferreras:1998th} (other noticeable degeneracies are the ones with the star formation history \cite{Gavazzi:2002} or with dust \cite{Pozzetti:2000kk}). However, the information encoded in the photometry and spectrum of a galaxy allows us to break (or mitigate in some cases) these degeneracies, and different methods have been introduced and exploited to provide a robust measurement. First of all, we need to underline here that in the cosmic chronometers approach we are interested in determining the relative age $dt$, and not the absolute age $t$; this means that we will not be dependent on any offset in the determination of the age, provided that the differential measurement is unbiased, and it has been shown that the relative ages are more robust and can achieve higher accuracy than absolute ones, removing many systematic effects when taking the differences \cite{Marin-Franch:2008cqi}. 

\begin{figure}[t!]
\includegraphics[width=0.99\textwidth]{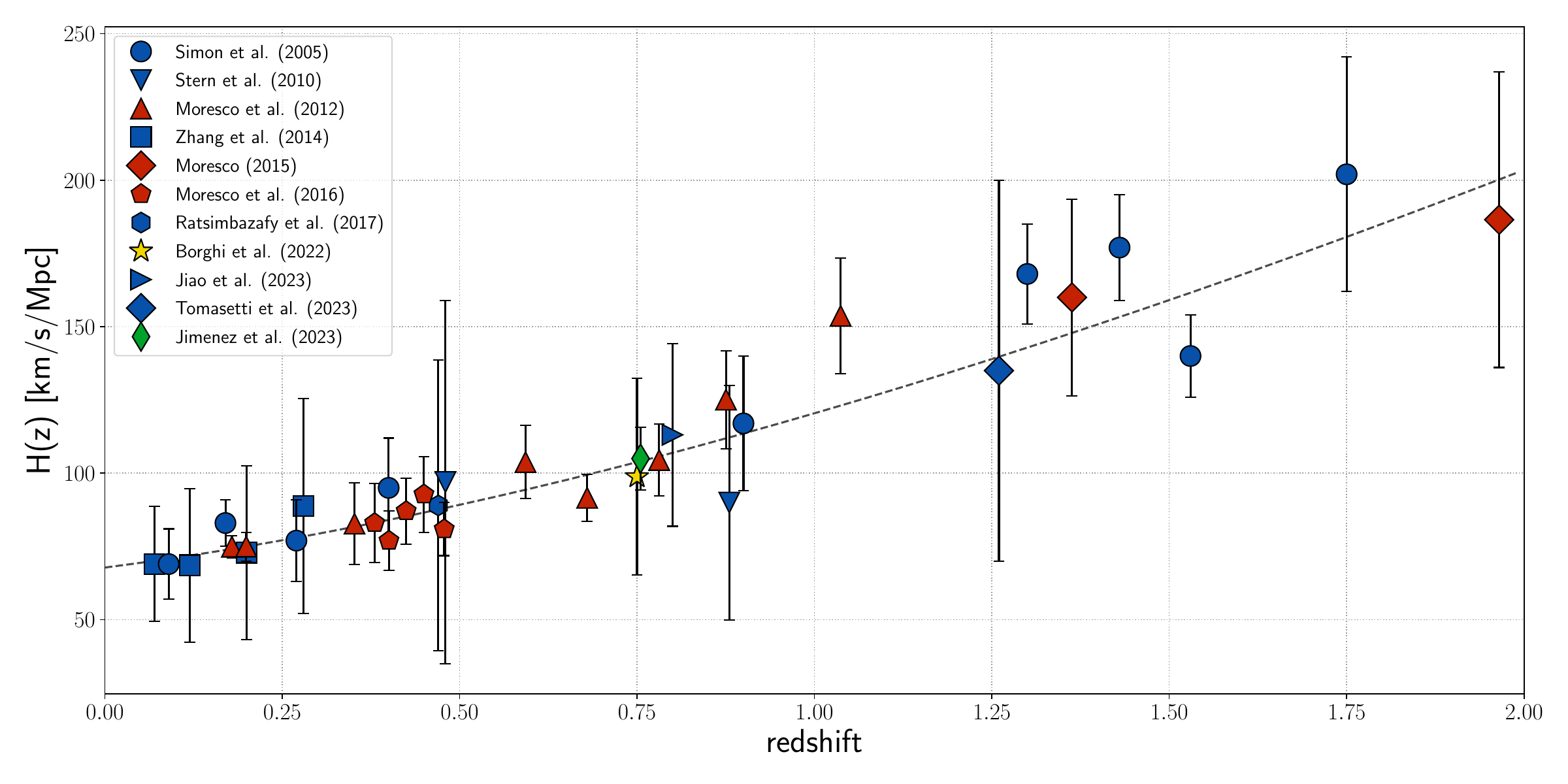}
\caption{Latest compilation of Hubble parameter measurements as a function of redshift obtained with cosmic chronometers. Different symbols indicate a measurement performed with a different method, as discussed in Sect.~\ref{sec:age} and presented in Tab.~\ref{tab:CC}.}
\label{fig:Hz}
\end{figure}

We summarize here in the following the main methods that have been used to derive $dt$, leaving to Ref.~\cite{Moresco:2022phi} a detailed discussion, and report in Tab.~\ref{tab:CC} and Fig.~\ref{fig:Hz} the Hubble parameters determined with the various approaches.\\

\begin{table}[t!]
\caption{Measurements of the Hubble parameter derived with the cosmic chronometers method obtained with the CC method (in units of \Hunit). The error on $H(z)$ reported here accounts only for the diagonal part of the covariance matrix, while for a proper analysis also the full covariance should be considered as discussed in Sect.~\ref{sec:age}. The method (M) used to derive the differential age $dt$ is also provided (full-spectrum fitting F, Lick indices L, $D4000$ D, Machine Learning ML), and the corresponding reference.}
\label{tab:CC}
\centering
\begin{tabular}{p{1cm}p{1cm}p{1cm}p{0.5cm}p{1.5cm}p{1cm}p{1cm}p{1cm}p{0.5cm}p{1.5cm}}
\hline\noalign{\smallskip}
$z$ & $H(z)$ & $\sigma_{H(z)}$ & M & reference & $z$ & $H(z)$ & $\sigma_{H(z)}$ & M & reference\\
\noalign{\smallskip}\svhline\noalign{\smallskip}
0.07 & 69.0 & 19.6 & F & \cite{Zhang:2012mp} & 0.593 & 104 & 13 & D & \cite{Moresco:2012jh}\\
0.09 & 69 & 12 & F & \cite{Simon:2004tf} & 0.68 & 92 & 8 & D & \cite{Moresco:2012jh}\\
0.12 & 68.6 & 26.2 & F & \cite{Zhang:2012mp} & 0.75 & 98.8 & 33.6 & L & \cite{Borghi:2021rft}$^*$\\
0.17 & 83 & 8 & F & \cite{Simon:2004tf} & 0.75 & 105 & 10.76 & ML & \cite{Jimenez:2023flo}$^*$\\
0.179 & 75 & 4 & D & \cite{Moresco:2012jh} & 0.781 & 105 & 12 & D & \cite{Moresco:2012jh}\\
0.199 & 75 & 5 & D & \cite{Moresco:2012jh} & 0.8 & 113.1 & 25.22 & F & \cite{Jiao:2022aep}$^*$\\
0.20 & 72.9 & 29.6 & F & \cite{Zhang:2012mp} & 0.875 & 125 & 17 & D & \cite{Moresco:2012jh}\\
0.27 & 77 & 14 & F & \cite{Simon:2004tf} & 0.88 & 90 & 40 & F & \cite{Stern:2009ep}\\
0.28 & 88.8 & 36.6 & F & \cite{Zhang:2012mp} & 0.9 &  117 &  23 & F & \cite{Simon:2004tf}\\
0.352 & 83 & 14 & D & \cite{Moresco:2012jh} & 1.037 & 154 & 20 & D & \cite{Moresco:2012jh}\\
0.38 & 83 & 13.5 & D & \cite{Moresco:2016mzx} & 1.26 & 135 & 65 & F & \cite{Tomasetti:2023kek}\\
0.4 & 95 & 17 & F & \cite{Simon:2004tf} & 1.3 & 168 & 17 & F & \cite{Simon:2004tf}\\
0.4004 & 77 & 10.2 & D & \cite{Moresco:2016mzx} & 1.363 & 160 & 33.6 & D & \cite{Moresco:2015cya}\\
0.425 & 87.1 & 11.2 & D & \cite{Moresco:2016mzx} & 1.43 & 177 & 18 & F & \cite{Simon:2004tf}\\
0.445 & 92.8 & 12.9 & D & \cite{Moresco:2016mzx} & 1.53 & 140 & 14 & F & \cite{Simon:2004tf}\\
0.47 & 89.0 & 49.6 & F & \cite{Ratsimbazafy:2017vga} & 1.75 & 202 & 40 & F & \cite{Simon:2004tf}\\
0.4783 & 80.9 & 9 & D & \cite{Moresco:2016mzx} & 1.965 & 186.5 & 50.4 & D & \cite{Moresco:2015cya}\\
0.48 & 97 & 62 & F & \cite{Stern:2009ep} & & & & & \\
\noalign{\smallskip}\hline\noalign{\smallskip}
\end{tabular}
{\scriptsize $^*$these data have been obtained from the same sample, and should not be used together in an analysis}
\end{table}

{\bf Full-spectral fitting.} The first approach that can be adopted is to exploit the full information available. Different codes are publicly available performing a fit to the photometric and spectroscopic data with theoretical spectra obtained combining the different physical ingredients of the stellar population of a galaxy (age, metallicity, mass, star formation history). This allows us to determine the ages of our chronometers, from which it is possible to derive the differential age by binning the data at different redshifts and estimating the quantity $dt$. Successful cosmic chronometers analyses have been performed with this method by several groups \cite{Simon:2004tf, Stern:2009ep, Zhang:2012mp, Ratsimbazafy:2017vga}, that have also demonstrated how with a good photometric and spectroscopic coverage the $dt$ measurement is very robust and independent of the specific assumptions chosen to model the spectra \cite{Jiao:2022aep, Tomasetti:2023kek}.

{\bf Analysis of absorption lines (Lick indices).} Instead of considering the full spectral information as in the previous method, an alternative is to study instead the absorption lines that characterize the spectra of passive galaxies, since they are known to be particularly sensitive to the age (e.g., Balmer lines), metal content (e.g., iron lines), and enhancement in $\alpha$ elements (e.g., magnesium lines) of the population. Similarly as before, it is possible to build theoretical models to interpret the strength of these features, known as Lick indices, as a function of the age, metallicity, and $\alpha$/Fe of the population, and therefore extract this information by analyzing a combination of indices. Also with this method, it has been recently obtained a determination of the Hubble parameter with the cosmic chronometer approach \cite{Borghi:2021rft}.\\

\begin{figure}[t!]
\includegraphics[width=0.99\textwidth]{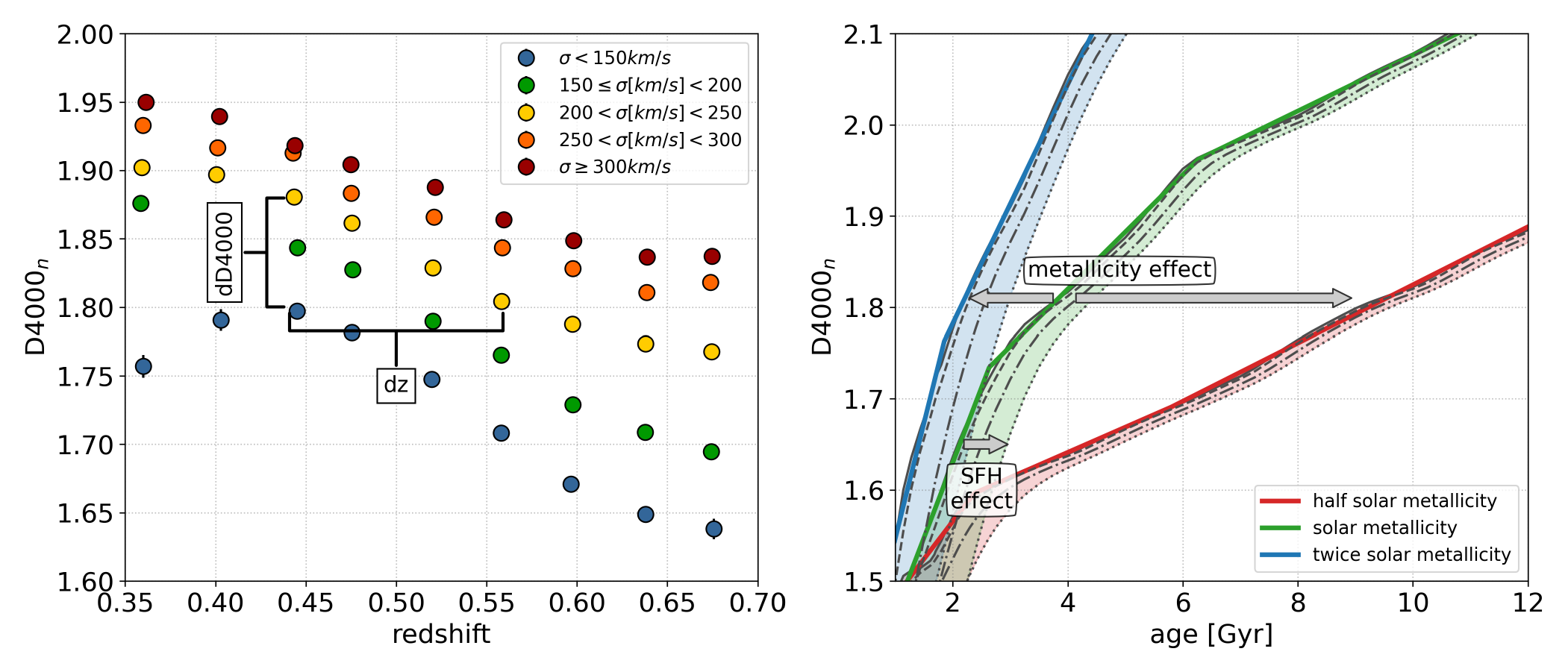}
\caption{Illustrative representation of how to apply the $D4000$ method in the cosmic chronometers approach (reproduced from \cite{Moresco:2022phi}. The left panel shows an example of median $D4000$-z values obtained from an analysis of SDSS BOSS survey, representing how to derive the quantity $dz/dD4000$. The right panel show for a specific model the $D4000$-age relations, where it is highlighted their dependence on metallicity (different colors) and star formation history (different lines). It is also possible to see how these relations can be approximated with piecewise linear relation with good accuracy.}
\label{fig:D4000z}
\end{figure}

{\bf Analysis of specific spectral features ($\mathbf{D4000}$).}
A slightly different approach consists in still considering spectroscopic features, but focusing on a specific spectral feature known to correlate particularly well with the age of the population instead of analyzing a joint combination of those. This approach, introduced by Ref.~\cite{Moresco:2010wh}, suggested using the break at 4000~\AA~restframe (known as $D4000$, see Fig.~\ref{fig:sel}) as an age indicator since it has been found that for cosmic chronometers it correlates linearly with the stellar age (at fixed metallicity and star formation history). If there is a linear relation between the $D4000$ and the age modeled as $D4000=A(Z, SFH)\times{\rm age}+B$, Ref.~\cite{Moresco:2012jh} proposed to rewrite Eq.~\ref{eq:CC} in the form:
\begin{svgraybox}
\begin{equation}
H(z) = -\frac{1}{1+z} A(SFR,Z/Z_{\odot})\frac{dz}{dD4000}\; ,
\end{equation}
\end{svgraybox}
where $A(SFR,Z/Z_{\odot})$ is a calibration parameter depending on the metallicity and star formation history obtained from stellar population synthesis models. We illustratively show how this method works in Fig.~\ref{fig:D4000z}. This equation improves on Eq.~\ref{eq:CC} since it allows to decouple statistical and systematic effects because the differential evolution will be measured on purely observational data ($dD4000$, purely including statistical errors), while the systematic errors will be all contained in the calibration parameter. To derive this properly, the metallicity of the sample has to be derived with independent measurements, or assumed in a realistic range; here we underline that given the cosmic chronometers selection described in Sect.~\ref{sec:sel}, we expect the metal content to be rather homogeneous as a function of redshift. However, it is fundamental to keep into account this effect, as well as the dependence of the results on the model to derive the calibration parameter, in the total covariance matrix, to properly include in the error budget all systematic effects as discussed below. With this approach, the most accurate $H(z)$ determinations of $H(z)$ have been derived to date, with accuracy going from $\sim$5\% at $z\sim0.15$, to $\sim$10\% at $z\sim0.8$, to $\sim$25\% at $z\sim2$ \cite{Moresco:2012jh, Moresco:2015cya, Moresco:2016mzx}.\\

{\bf Other approaches.} Recently, Ref.~\cite{Jimenez:2023flo} demonstrated that it is possible to apply the cosmic chronometer approach also considering purely photometric data, using a Machine Learning algorithm trained on well-determined aged (obtained from Lick indices analysis). This method, while needing to rely on larger samples ($O(10^4)$) to determine the upper envelope of the age-redshift relation, is extremely promising in view of all the surveys that will provide photometry for almost the full sky in the near future (e.g. Euclid \cite{EUCLID:2011zbd} and Rubin \cite{LSST:2008ijt}).\\

{\bf Cosmic chronometers full covariance matrix.} The full covariance matrix for cosmic chronometers has been formalized in Ref.~\cite{Moresco:2020fbm}, and can be divided into different contributions:
\begin{eqnarray}
{\rm Cov}_{ij}^{\rm tot}&=& {\rm Cov}_{ij}^{\rm stat}+ \nonumber\\
& &+ {\rm Cov}_{ij}^{\rm met} + {\rm Cov}_{ij}^{\rm young}+\nonumber\\
& &+ {\rm Cov}_{ij}^{\rm SFH}+{\rm Cov}_{ij}^{\rm IMF}+{\rm Cov}_{ij}^{\rm st. lib.}+{\rm Cov}_{ij}^{\rm SPS} \; .
\label{eq:cov}
\end{eqnarray}
The first row of Eq.~\ref{eq:cov} reports the component related to the statistical errors (e.g. the one related to the measurement of $dD4000$ if using the $D4000$ method to measure $dt$). The second row reports the systematic contribution to the covariance due to uncertainty in the determination of the metallicity of the sample, or to a non-perfect selection of the sample when residual contamination by a young population is not completely removed; this second effect has been quantified in Ref.~\cite{Moresco:2018xdr}, where a recipe to derive it directly from observable is provided. These two effects are expected to provide diagonal contributions to the covariance since they depend on uncorrelated properties of the galaxies selected. The last row describes instead the systematic component of the covariance due to the ingredients of the model assumed. It has been divided into various parts, i.e. the one related to the uncertainty on the stellar population synthesis model considered ({\it SPS}), the one related to the stellar library that the model adopts ({\it st.lib.}), the one related to the initial mass function considered in the model ({\it IMF}), and the one depending on the assumed star formation history ({\it SFH}). All these parts have been thoroughly explored in Ref.~\cite{Moresco:2020fbm}, where also a Jupyter notebook is provided as an example of how to derive the total covariance matrix for cosmic chronometers.

To conclude, we note here that several other effects have been considered that might in principle affect the cosmological measurements, such as the progenitor bias or a dependence on the stellar mass of the sample. These effects, discussed in Ref.~\cite{Moresco:2022phi}, are expected however not to have a significant impact on the analysis due to the strict selection process described in Sect.~\ref{sec:sel} and since in the differential approach relative ages are estimated in homogeneous samples really close in cosmic time, minimizing therefore by definition these effects.

\section{Cosmic chronometers and the Hubble constant}
\label{sec:H0}

\begin{table}[t!]
\caption{Cosmological parameter constraints that can be obtained with present-day and future CC measurements (as presented in Sect.~\ref{sec:H0}) in two different cosmologies, a flat $\Lambda$CDM cosmology (upper rows) and a more general open $w$CDM (lower rows). The Hubble constant \Ho~is given in units \Hunit.}
\label{tab:H0}
\centering
\begin{tabular}{p{2cm}p{1.25cm}p{0.9cm}p{1.48cm}p{0.9cm}p{1.25cm}p{0.9cm}p{1.25cm}p{0.9cm}}
\hline\noalign{\smallskip}
 & \Ho & \% acc & \omegam & \% acc & \omegade & \% acc & $w$ & \% acc \\ 
\noalign{\smallskip}\svhline\noalign{\smallskip}
 & \multicolumn{8}{c}{flat $\Lambda$CDM} \\
\cline{2-9}\noalign{\smallskip}
current dataset & $66.7\pm5.3$ & 8.\% & $0.33^{+0.08}_{-0.06}$ & 20.6\% & -- & -- & -- & --\\
combined & $69.0\pm2.1$ & 3\% & $0.3\pm0.01$ & 4.4\% & -- & -- & -- & --\\
optimistic & $68.2\pm0.7$ & 1\% & $0.29\pm 0.01$ & 4.2\% & -- & -- & -- & --\\
\hline\noalign{\smallskip}
 & \multicolumn{8}{c}{open $w$CDM} \\
\cline{2-9}\noalign{\smallskip}
current dataset & $68.0^{+8.8}_{-7.1}$ & 11.7\% & $0.22^{+0.15}_{-0.12}$ & 62\% & $0.51^{+0.24}_{-0.26}$ & 48\% & $-1.6^{+0.8}_{-0.9}$ & 52\%\\
combined & $71.6^{+3.1}_{-2.6}$ & 4\% & $0.29^{+0.09}_{-0.09}$ & 31\% & $0.67^{+0.18}_{-0.14}$ & 24\% & $-1.2^{+0.3}_{ -0.4}$ & 29\%\\
optimistic & $70.8^{+2.8}_{-1.9}$ & 3.3\% & $0.3^{+0.08}_{-0.09}$ & 28\% & $0.68^{+0.18}_{-0.11}$ & 22\% & $-1.2^{+0.3}_{-0.4}$ & 29\%\\
\noalign{\smallskip}\hline\noalign{\smallskip}
\end{tabular}
\end{table}

As shown in Fig.~\ref{fig:Hz}, cosmic chronometers do not directly constrain the Hubble constant, but rather measure the Hubble parameter independently of any cosmological assumptions. For this reason, they represent the ideal testbed to compare different cosmological models.
While not measuring \Ho~directly, however, it can be derived from these data in different ways. 

A possible solution is to estimate the Hubble constant as the extrapolated value of $H(z)$ at $z=0$. To exploit at most the cosmology-independent nature of these measurements, different methods have been suggested, either based on Gaussian Process regression (e.g. \cite{Seikel:2012, Haridasu:2018gqm, Gomez-Valent:2018hwc}, but see also \cite{OColgain:2021pyh} on the assumptions of the method and potential dependence of the measured errors on the assumed kernel) or on Machine Learning (e.g. \cite{Arjona:2019fwb, Mehrabi:2021cob}). The most recent constraint obtained with these methods give $H_0=70.7\pm6.7$ \Hunit~when considering the full covariance matrix \cite{Favale:2023lnp}.

\begin{figure}[t!]
\includegraphics[width=0.5\textwidth]{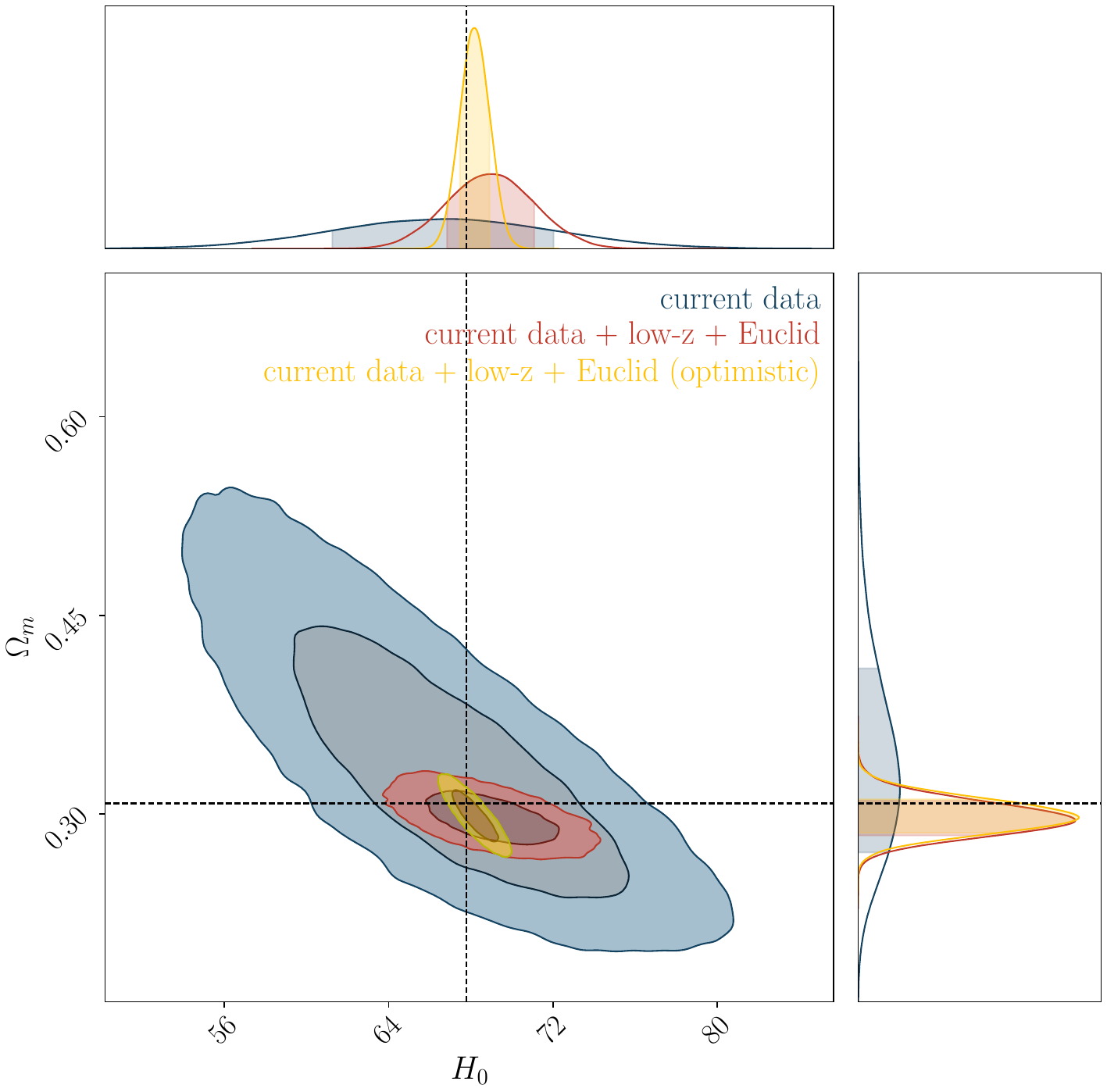}
\includegraphics[width=0.5\textwidth]{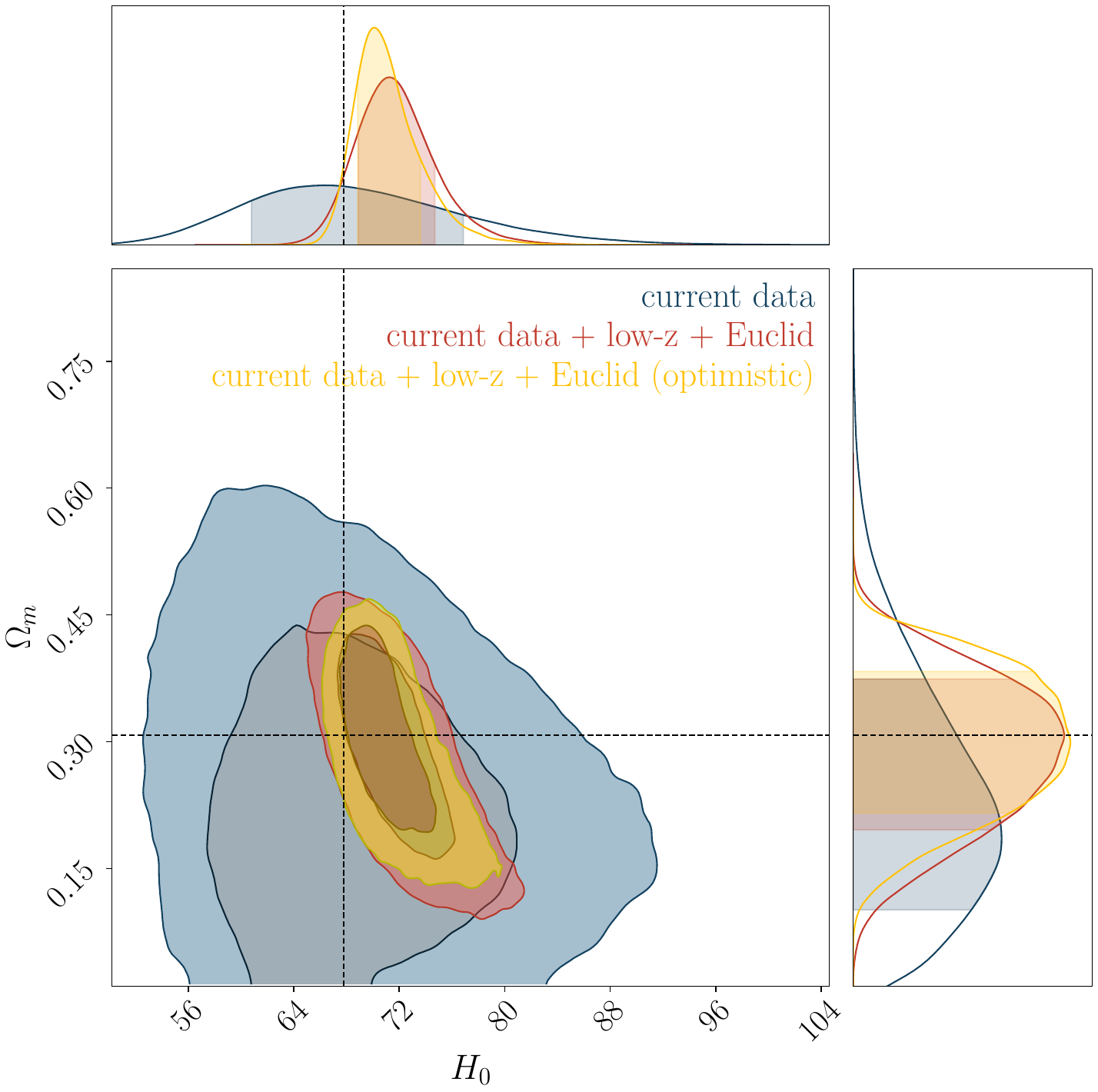}
\caption{Constraints in the \Ho-\omegam~plane obtained from the analysis of cosmic chronometers for two cosmological models, a flat $\Lambda$CDM (left panel) and an open wCDM (right panel). The different colors refer to the analysis of the current dataset presented in Tab.~\ref{tab:CC} (in blue), and with the forecasts for future measurements in two configurations (a standard one in red, and an optimistic one in gold, as discussed in Sect.~\ref{sec:H0}. Dashed lines report as a reference the cosmological parameters obtained by Planck2018 \cite{Planck:2018vyg}.}
\label{fig:cosmo}
\end{figure}

The most direct way to derive \Ho~is to assume a cosmological model and fit the data presented in Tab.~\ref{tab:CC}, properly considering their covariance matrix as discussed above. We assumed here two cosmological models, a flat $\Lambda$CDM model(the one with the least free parameters, \Ho~and \omegam) and the open wCDM model, where also the curvature of the Universe and the dark energy equation of state parameter are left free. This can give us a range of the expected constraints while varying the number of parameters in the fit. Moreover, we also extend our analysis by adding to the current $H(z)$ measurements the forecasts discussed by Ref.~\cite{Moresco:2022phi}, where two sets of simulated data are provided as expected by the analysis of the just launched Euclid mission and from the exploitation of newly derived data (such as the latest SDSS BOSS and eBOSS surveys). In this case, two different simulations are explored, a standard one where current data are combined with these simulated data ({\it combined}) and one where a more optimistic impact of systematic error is considered for the simulated data, minimizing the contribution due to different models in the analysis. The constraints are shown in Fig.~\ref{fig:cosmo}, and the results are reported in Tab.~\ref{tab:H0}.
We find that current data are able to measure \Ho with an accuracy of 8\% in a flat $\lambda$CDM cosmology, and of 11\% in an open wCDM cosmology, with values $H_0=66.7\pm5.3$ and $68.0^{+8.8}_{-7.1}$ \Hunit, respectively. As can be seen, the constraints at the moment are preferring a lower value of $H_0$, confirming the results obtained with Gaussian Processes, but with an errorbar that does not allows us to significantly discriminate between SH0ES and CMB values. However, it is interesting to notice how future data could significantly improve, reaching an accuracy of around 3--1\% depending on the cosmological model assumed and on the simulation considered. This sheds a very promising light on this method since it suggests that future analysis and a more thorough study of the models impacting on the systematic errors might provide competitive \Ho~measurements, that can help in giving important independent pieces of evidence to address the \Ho~tension.

Finally, another possibility is to explore how much the constraints would improve by combining cosmic chronometers with other cosmological probes. This has been done by several authors (e.g. \cite{Moresco:2012by, Moresco:2016nqq, Yu:2017iju, Zhao:2017cud, Bonilla:2020wbn}), and the most recent measurement to date obtained combining SNeIa, BAO, cosmic chronometers and Gamma-Ray Bursts data gives $H_0=67.2^{+3.4}_{-3.2}$ \Hunit (for a $\Lambda$CDM cosmology, $H_0=66.9^{+3.5}_{-3.4}$ \Hunit~for an open wCDM cosmology, \cite{Cogato:2023atm}).


\begin{acknowledgement}
MM acknowledges the grants ASI n.I/023/12/0 and ASI n.2018-23-HH.0., and support from MIUR, PRIN 2017 (grant 20179ZF5KS).
\end{acknowledgement}




\begin{thebibliography}{99}

\bibitem{Abdalla:2022yfr}
E.~Abdalla, G.~Franco Abell\'an, A.~Aboubrahim, A.~Agnello, O.~Akarsu, Y.~Akrami, G.~Alestas, D.~Aloni, L.~Amendola and L.~A.~Anchordoqui, \textit{et al.}
JHEAp \textbf{34}, 49-211 (2022)
doi:10.1016/j.jheap.2022.04.002
[arXiv:2203.06142 [astro-ph.CO]].

\bibitem{Arnouts:2013kba}
S.~Arnouts, E.~Le Floc'h, J.~Chevallard, B.~D.~Johnson, O.~Ilbert, M.~Treyer, H.~Aussel, P.~Capak, D.~B.~Sanders and N.~Scoville, \textit{et al.}
Astron. Astrophys. \textbf{558}, A67 (2013)
doi:10.1051/0004-6361/201321768
[arXiv:1309.0008 [astro-ph.CO]].

\bibitem{Arjona:2019fwb}
R.~Arjona and S.~Nesseris,
Phys. Rev. D \textbf{101}, no.12, 123525 (2020)
doi:10.1103/PhysRevD.101.123525
[arXiv:1910.01529 [astro-ph.CO]].

\bibitem{Bonilla:2020wbn}
A.~Bonilla, S.~Kumar and R.~C.~Nunes,
Eur. Phys. J. C \textbf{81}, no.2, 127 (2021)
doi:10.1140/epjc/s10052-021-08925-z
[arXiv:2011.07140 [astro-ph.CO]].
\bibitem{Borghi:2021zsr}
N.~Borghi, M.~Moresco, A.~Cimatti, A.~Huchet, S.~Quai and L.~Pozzetti,
Astrophys. J. \textbf{927}, no.2, 164 (2022)
doi:10.3847/1538-4357/ac3240
[arXiv:2106.14894 [astro-ph.GA]].

\bibitem{Borghi:2021rft}
N.~Borghi, M.~Moresco and A.~Cimatti,
Astrophys. J. Lett. \textbf{928}, no.1, L4 (2022)
doi:10.3847/2041-8213/ac3fb2
[arXiv:2110.04304 [astro-ph.CO]].

\bibitem{Brout:2022vxf}
D.~Brout, D.~Scolnic, B.~Popovic, A.~G.~Riess, J.~Zuntz, R.~Kessler, A.~Carr, T.~M.~Davis, S.~Hinton and D.~Jones, \textit{et al.}
Astrophys. J. \textbf{938}, no.2, 110 (2022)
doi:10.3847/1538-4357/ac8e04
[arXiv:2202.04077 [astro-ph.CO]].

\bibitem{Caputi:2012uf}
K.~I.~Caputi, J.~S.~Dunlop, R.~J.~McLure, J.~Huang, G.~G.~Fazio, M.~L.~N.~Ashby, M.~Castellano, A.~Fontana, M.~Cirasuolo and O.~Almaini, \textit{et al.}
Astrophys. J. Lett. \textbf{750}, L20 (2012)
doi:10.1088/2041-8205/750/1/L20
[arXiv:1202.0496 [astro-ph.CO]].

\bibitem{Carnall:2018}
A.C.~Carnall, R.J.~McLure, J.S.~Dunlop, R.~Dav{\'e}
Mon. Not. Roy. Astron. Soc. \textbf{480}, 4379-4401 (2018)
doi:10.1093/mnras/sty2169
[arXiv:astro-ph/1712.04452 [astro-ph]].

\bibitem{Carnall:2019}
A.C.~Carnall, R.J.~McLure, J.S.~Dunlop, F.~Cullen, D.J.~McLeod, \textit{et al.}
Mon. Not. Roy. Astron. Soc. \textbf{490}, 417-439 (2019)
doi:10.1093/mnras/stz2544
[arXiv:astro-ph/1903.11082 [astro-ph]].

\bibitem{Carnall:2023}
A.C.~Carnall, D.J.~McLeod, R.J.~McLure, J.S.~Dunlop, R.~Begley, F.~Cullen, \textit{et al.}
Mon. Not. Roy. Astron. Soc. \textbf{520}, 3974-3985 (2023)
doi:10.1093/mnras/stad369
[arXiv:astro-ph/2208.00986 [astro-ph]].

\bibitem{Choi:2014vfa}
J.~Choi, C.~Conroy, J.~Moustakas, G.~J.~Graves, B.~P.~Holden, M.~Brodwin, M.~J.~I.~Brown and P.~G.~van Dokkum,
Astrophys. J. \textbf{792}, no.2, 95 (2014)
doi:10.1088/0004-637X/792/2/95
[arXiv:1403.4932 [astro-ph.GA]].

\bibitem{Cimatti:2004gq}
A.~Cimatti, E.~Daddi, A.~Renzini, P.~Cassata, E.~Vanzella, L.~Pozzetti, S.~Cristiani, A.~Fontana, G.~Rodighiero and M.~Mignoli, \textit{et al.}
Nature \textbf{430}, 184-187 (2004)
doi:10.1038/nature02668
[arXiv:astro-ph/0407131 [astro-ph]].

\bibitem{Cimatti:2023gil}
A.~Cimatti and M.~Moresco,
[arXiv:2302.07899 [astro-ph.CO]].

\bibitem{Citro:2016}
A.~Citro, L.~Pozzetti, M.~Moresco, A.~Cimatti
Astron. Astrophys. \textbf{592}, A19 (2016)
doi:10.1051/0004-6361/201527772
[arXiv:1604.07826 [astro-ph.CO]].

\bibitem{Cogato:2023atm}
F.~Cogato, M.~Moresco, L.~Amati and A.~Cimatti,
[arXiv:2309.01375 [astro-ph.CO]].

\bibitem{OColgain:2021pyh}
E.~\'O Colg\'ain and M.~M.~Sheikh-Jabbari,
Eur. Phys. J. C \textbf{81}, no.10, 892 (2021)
doi:10.1140/epjc/s10052-021-09708-2
[arXiv:2101.08565 [astro-ph.CO]].

\bibitem{Conroy:2013iaa}
C.~Conroy, G.~Graves and P.~van Dokkum,
Astrophys. J. \textbf{780}, 33 (2014)
doi:10.1088/0004-637X/780/1/33
[arXiv:1303.6629 [astro-ph.CO]].

\bibitem{Daddi:2003hu}
E.~Daddi, A.~Cimatti, A.~Renzini, J.~Vernet, C.~Conselice, L.~Pozzetti, M.~Mignoli, P.~Tozzi, T.~J.~Broadhurst and S.~di Serego Alighieri, \textit{et al.}
Astrophys. J. Lett. \textbf{600}, L127-L130 (2004)
doi:10.1086/381020
[arXiv:astro-ph/0308456 [astro-ph]].

\bibitem{Daddi:2005uv}
E.~Daddi, A.~Renzini, N.~Pirzkal, A.~Cimatti, S.~Malhotra, M.~Stiavelli, C.~Xu, A.~Pasquali, J.~E.~Rhoads and M.~Brusa, \textit{et al.}
Astrophys. J. \textbf{626}, 680-697 (2005)
doi:10.1086/430104
[arXiv:astro-ph/0503102 [astro-ph]].

\bibitem{Estrada-Carpenter:2019}
V.~Estrada-Carpenter, C.~Papovich, I.~Momcheva, G.~Brammer, J.~Long, R.F.~Quadri, J.~Bridge, M.~Dickinson, H.~Ferguson, S.~Finkelstein, \textit{et al.}
Astrophys. J. \textbf{870}, 133-160 (2019)
doi:10.3847/1538-4357/aaf22e
[arXiv:astro-ph/1810.02824 [astro-ph]].

\bibitem{Favale:2023lnp}
A.~Favale, A.~G\'omez-Valent and M.~Migliaccio,
Mon. Not. Roy. Astron. Soc. \textbf{523}, no.3, 3406-3422 (2023)
doi:10.1093/mnras/stad1621
[arXiv:2301.09591 [astro-ph.CO]].

\bibitem{Ferreras:1998th}
I.~Ferreras, S.~Charlot and J.~Silk,
Astrophys. J. \textbf{521}, 81 (1999)
doi:10.1086/307513
[arXiv:astro-ph/9803235 [astro-ph]].

\bibitem{Fontana:2006xg}
A.~Fontana, S.~Salimbeni, A.~Grazian, E.~Giallongo, L.~Pentericci, M.~Nonino, F.~Fontanot, N.~Menci, P.~Monaco and S.~Cristiani, \textit{et al.}
Astron. Astrophys. \textbf{459}, 745-757 (2006)
doi:10.1051/0004-6361:20065475
[arXiv:astro-ph/0609068 [astro-ph]].

\bibitem{Gallazzi:2005df}
A.~Gallazzi, S.~Charlot, J.~Brinchmann, S.~D.~M.~White and C.~A.~Tremonti,
Mon. Not. Roy. Astron. Soc. \textbf{362}, 41-58 (2005)
doi:10.1111/j.1365-2966.2005.09321.x
[arXiv:astro-ph/0506539 [astro-ph]].

\bibitem{Gallazzi:2014rca}
A.~Gallazzi, E.~F.~Bell, S.~Zibetti, J.~Brinchmann and D.~D.~Kelson,
Astrophys. J. \textbf{788}, 72 (2014)
doi:10.1088/0004-637X/788/1/72
[arXiv:1404.5624 [astro-ph.GA]].

\bibitem{Gavazzi:2002}
G.~Gavazzi, C.~Bonfanti, G.~Sanvito, A.~Boselli and M.~Scodeggio,
Astrophys. J. \textbf{576}, 1 (2002)
doi:10.1086/341730

\bibitem{Gomez-Valent:2018hwc}
A.~G\'omez-Valent and L.~Amendola,
JCAP \textbf{04}, 051 (2018)
doi:10.1088/1475-7516/2018/04/051
[arXiv:1802.01505 [astro-ph.CO]].

\bibitem{Haridasu:2018gqm}
B.~S.~Haridasu, V.~V.~Lukovi\'c, M.~Moresco and N.~Vittorio,
JCAP \textbf{10}, 015 (2018)
doi:10.1088/1475-7516/2018/10/015
[arXiv:1805.03595 [astro-ph.CO]].

\bibitem{Heavens:2004sr}
A.~Heavens, B.~Panter, R.~Jimenez and J.~Dunlop,
Nature \textbf{428}, 625 (2004)
doi:10.1038/nature02474
[arXiv:astro-ph/0403293 [astro-ph]].

\bibitem{Huterer:2017buf}
D.~Huterer and D.~L.~Shafer,
Rept. Prog. Phys. \textbf{81}, no.1, 016901 (2018)
doi:10.1088/1361-6633/aa997e
[arXiv:1709.01091 [astro-ph.CO]].

\bibitem{Ilbert:2008hz}
O.~Ilbert, P.~Capak, M.~Salvato, H.~Aussel, H.~J.~McCracken, D.~B.~Sanders, N.~Scoville, J.~Kartaltepe, S.~Arnouts and E.~L.~Floc'h, \textit{et al.}
Astrophys. J. \textbf{690}, 1236-1249 (2009)
doi:10.1088/0004-637X/690/2/1236
[arXiv:0809.2101 [astro-ph]].

\bibitem{Ilbert:2013bf}
O.~Ilbert, H.~J.~McCracken, O.~L.~Fevre, P.~Capak, J.~Dunlop, S.~Arnouts, H.~Aussel, K.~Caputi, J.~Comparat and Q.~Guo, \textit{et al.}
Astron. Astrophys. \textbf{556}, A55 (2013)
doi:10.1051/0004-6361/201321100
[arXiv:1301.3157 [astro-ph.CO]].

\bibitem{LSST:2008ijt}
\v{Z}.~Ivezi\'c \textit{et al.} [LSST],
Astrophys. J. \textbf{873}, no.2, 111 (2019)
doi:10.3847/1538-4357/ab042c
[arXiv:0805.2366 [astro-ph]].

\bibitem{Jiao:2022aep}
K.~Jiao, N.~Borghi, M.~Moresco and T.~J.~Zhang,
Astrophys. J. Suppl. \textbf{265}, no.2, 48 (2023)
doi:10.3847/1538-4365/acbc77
[arXiv:2205.05701 [astro-ph.CO]].

\bibitem{Jimenez:1996at}
R.~Jimenez, P.~Thejll, U.~Jorgensen, J.~MacDonald and B.~Pagel,
Mon. Not. Roy. Astron. Soc. \textbf{282}, 926-942 (1996)
doi:10.1093/mnras/282.3.926
[arXiv:astro-ph/9602132 [astro-ph]].

\bibitem{Jimenez:2001gg}
R.~Jimenez and A.~Loeb,
Astrophys. J. \textbf{573}, 37-42 (2002)
doi:10.1086/340549
[arXiv:astro-ph/0106145 [astro-ph]].

\bibitem{Jimenez:2003iv}
R.~Jimenez, L.~Verde, T.~Treu and D.~Stern,
Astrophys. J. \textbf{593}, 622-629 (2003)
doi:10.1086/376595
[arXiv:astro-ph/0302560 [astro-ph]].

\bibitem{Jimenez:2019onw}
R.~Jimenez, A.~Cimatti, L.~Verde, M.~Moresco and B.~Wandelt,
JCAP \textbf{03}, 043 (2019)
doi:10.1088/1475-7516/2019/03/043
[arXiv:1902.07081 [astro-ph.CO]].

\bibitem{Jimenez:2023flo}
R.~Jimenez, M.~Moresco, L.~Verde and B.~D.~Wandelt,
[arXiv:2306.11425 [astro-ph.CO]].

\bibitem{Hubble:1929ig}
E.~Hubble,
Proc. Nat. Acad. Sci. \textbf{15}, 168-173 (1929)
doi:10.1073/pnas.15.3.168

\bibitem{Krauss:2003em}
L.~M.~Krauss and B.~Chaboyer,
Science \textbf{299}, 65-70 (2003)
doi:10.1126/science.1075631

\bibitem{Kriek:2019}
M.~Kriek, S.H.~Price, C.~Conroy, K.A.~Suess, L.~Mowla, I.~Pasha, R.~Bezanson, P.~Van~Dokkum, G.~Barro,
Astrophys. J. Lett. \textbf{880}, no.2, L31 (2019)
doi:10.3847/2041-8213/ab2e75
[arXiv:astro-ph/1907.04327 [astro-ph]].

\bibitem{EUCLID:2011zbd}
R.~Laureijs \textit{et al.} [EUCLID],
[arXiv:1110.3193 [astro-ph.CO]].

\bibitem{Marin-Franch:2008cqi}
A.~Marin-Franch, A.~Aparicio, G.~Piotto, A.~Rosenberg, B.~Chaboyer, A.~Sarajedini, M.~Siegel, J.~Anderson, L.~R.~Bedin and A.~Dotter, \textit{et al.}
Astrophys. J. \textbf{694}, 1498-1516 (2009)
doi:10.1088/0004-637X/694/2/1498
[arXiv:0812.4541 [astro-ph]].

\bibitem{McDermid:2015ska}
R.~M.~McDermid, K.~Alatalo, L.~Blitz, F.~Bournaud, M.~Bureau, M.~Cappellari, A.~F.~Crocker, R.~L.~Davies, T.~A.~Davis and P.~T.~de Zeeuw, \textit{et al.}
Mon. Not. Roy. Astron. Soc. \textbf{448}, no.4, 3484-3513 (2015)
doi:10.1093/mnras/stv105
[arXiv:1501.03723 [astro-ph.GA]].

\bibitem{Mehrabi:2021cob}
A.~Mehrabi and M.~Rezaei,
Astrophys. J. \textbf{923}, no.2, 274 (2021)
doi:10.3847/1538-4357/ac2fff
[arXiv:2110.14950 [astro-ph.CO]].

\bibitem{Merlin:2019}
E.~Merlin, F.~Fortuni, M.~Torelli, P.~Santini, M.~Castellano, A.~Fontana, A.~Grazian, L.~Pentericci, S.~Pilo, K.B.~Schmidt
Mon. Not. Roy. Astron. Soc. \textbf{490}, no.3, 3309-3328 (2019)
doi:10.1093/mnras/stz2615
[arXiv:1909.07996 [astro-ph.GA]].

\bibitem{Mignoli:2008yz}
M.~Mignoli, G.~Zamorani, M.~Scodeggio, A.~Cimatti, C.~Halliday, S.~J.~Lilly, L.~Pozzetti, D.~Vergani, C.~M.~Carollo and T.~Contini, \textit{et al.}
Astron. Astrophys. \textbf{493}, 39 (2009)
doi:10.1051/0004-6361:200810520
[arXiv:0810.2245 [astro-ph]].

\bibitem{Moresco:2010wh}
M.~Moresco, R.~Jimenez, A.~Cimatti and L.~Pozzetti,
JCAP \textbf{03}, 045 (2011)
doi:10.1088/1475-7516/2011/03/045
[arXiv:1010.0831 [astro-ph.CO]].

\bibitem{Moresco:2012jh}
M.~Moresco, A.~Cimatti, R.~Jimenez, L.~Pozzetti, G.~Zamorani, M.~Bolzonella, J.~Dunlop, F.~Lamareille, M.~Mignoli and H.~Pearce, \textit{et al.}
JCAP \textbf{08}, 006 (2012)
doi:10.1088/1475-7516/2012/08/006
[arXiv:1201.3609 [astro-ph.CO]].

\bibitem{Moresco:2012by}
M.~Moresco, L.~Verde, L.~Pozzetti, R.~Jimenez and A.~Cimatti,
JCAP \textbf{07}, 053 (2012)
doi:10.1088/1475-7516/2012/07/053
[arXiv:1201.6658 [astro-ph.CO]].

\bibitem{Moresco:2013wsa}
M.~Moresco, L.~Pozzetti, A.~Cimatti, G.~Zamorani, M.~Bolzonella, F.~Lamareille, M.~Mignoli, E.~Zucca, S.~J.~Lilly and C.~M.~Carollo, \textit{et al.}
Astron. Astrophys. \textbf{558}, A61 (2013)
doi:10.1051/0004-6361/201321797
[arXiv:1305.1308 [astro-ph.CO]].

\bibitem{Moresco:2015cya}
M.~Moresco,
Mon. Not. Roy. Astron. Soc. \textbf{450}, no.1, L16-L20 (2015)
doi:10.1093/mnrasl/slv037
[arXiv:1503.01116 [astro-ph.CO]].

\bibitem{Moresco:2016mzx}
M.~Moresco, L.~Pozzetti, A.~Cimatti, R.~Jimenez, C.~Maraston, L.~Verde, D.~Thomas, A.~Citro, R.~Tojeiro and D.~Wilkinson,
JCAP \textbf{05}, 014 (2016)
doi:10.1088/1475-7516/2016/05/014
[arXiv:1601.01701 [astro-ph.CO]].

\bibitem{Moresco:2016nqq}
M.~Moresco, R.~Jimenez, L.~Verde, A.~Cimatti, L.~Pozzetti, C.~Maraston and D.~Thomas,
JCAP \textbf{12}, 039 (2016)
doi:10.1088/1475-7516/2016/12/039
[arXiv:1604.00183 [astro-ph.CO]].

\bibitem{Moresco:2017hwt}
M.~Moresco and F.~Marulli,
Mon. Not. Roy. Astron. Soc. \textbf{471}, no.1, L82-L86 (2017)
doi:10.1093/mnrasl/slx112
[arXiv:1705.07903 [astro-ph.CO]].

\bibitem{Moresco:2018xdr}
M.~Moresco, R.~Jimenez, L.~Verde, L.~Pozzetti, A.~Cimatti and A.~Citro,
Astrophys. J. \textbf{868}, no.2, 84 (2018)
doi:10.3847/1538-4357/aae829
[arXiv:1804.05864 [astro-ph.CO]].

\bibitem{Moresco:2020fbm}
M.~Moresco, R.~Jimenez, L.~Verde, A.~Cimatti and L.~Pozzetti,
Astrophys. J. \textbf{898}, no.1, 82 (2020)
doi:10.3847/1538-4357/ab9eb0
[arXiv:2003.07362 [astro-ph.GA]].

\bibitem{Moresco:2022phi}
M.~Moresco, L.~Amati, L.~Amendola, S.~Birrer, J.~P.~Blakeslee, M.~Cantiello, A.~Cimatti, J.~Darling, M.~Della Valle and M.~Fishbach, \textit{et al.}
Living Rev. Rel. \textbf{25}, no.1, 6 (2022)
doi:10.1007/s41114-022-00040-z
[arXiv:2201.07241 [astro-ph.CO]].

\bibitem{Muzzin:2013yxa}
A.~Muzzin, D.~Marchesini, M.~Stefanon, M.~Franx, H.~J.~McCracken, B.~Milvang-Jensen, J.~S.~Dunlop, J.~P.~U.~Fynbo, O.~Le Fevre and G.~Brammer, \textit{et al.}
Astrophys. J. \textbf{777}, 18 (2013)
doi:10.1088/0004-637X/777/1/18
[arXiv:1303.4409 [astro-ph.CO]].

\bibitem{Onodera:2012en}
M.~Onodera, A.~Renzini, M.~Carollo, M.~Cappellari, C.~Mancini, V.~Strazzullo, E.~Daddi, N.~Arimoto, R.~Gobat and Y.~Yamada, \textit{et al.}
Astrophys. J. \textbf{755}, 26 (2012)
doi:10.1088/0004-637X/755/1/26
[arXiv:1206.1540 [astro-ph.CO]].

\bibitem{Onodera:2014rpa}
M.~Onodera, C.~M.~Carollo, A.~Renzini, M.~Cappellari, C.~Mancini, N.~Arimoto, E.~Daddi, R.~Gobat, V.~Strazzullo and S.~Tacchella, \textit{et al.}
Astrophys. J. \textbf{808}, no.2, 161 (2015)
doi:10.1088/0004-637X/808/2/161
[arXiv:1411.5023 [astro-ph.GA]].

\bibitem{Planck:2018vyg}
N.~Aghanim \textit{et al.} [Planck],
Astron. Astrophys. \textbf{641}, A6 (2020)
[erratum: Astron. Astrophys. \textbf{652}, C4 (2021)]
doi:10.1051/0004-6361/201833910
[arXiv:1807.06209 [astro-ph.CO]].

\bibitem{SupernovaCosmologyProject:1998vns}
S.~Perlmutter \textit{et al.} [Supernova Cosmology Project],
Astrophys. J. \textbf{517}, 565-586 (1999)
doi:10.1086/307221
[arXiv:astro-ph/9812133 [astro-ph]].

\bibitem{Pozzetti:2000kk}
L.~Pozzetti and F.~Mannucci,
Mon. Not. Roy. Astron. Soc. \textbf{317}, L17 (2000)
doi:10.1046/j.1365-8711.2000.03829.x
[arXiv:astro-ph/0006430 [astro-ph]].

\bibitem{Ratsimbazafy:2017vga}
A.~L.~Ratsimbazafy, S.~I.~Loubser, S.~M.~Crawford, C.~M.~Cress, B.~A.~Bassett, R.~C.~Nichol and P.~V\"ais\"anen,
Mon. Not. Roy. Astron. Soc. \textbf{467}, no.3, 3239-3254 (2017)
doi:10.1093/mnras/stx301
[arXiv:1702.00418 [astro-ph.CO]].

\bibitem{SupernovaSearchTeam:1998fmf}
A.~G.~Riess \textit{et al.} [Supernova Search Team],
Astron. J. \textbf{116}, 1009-1038 (1998)
doi:10.1086/300499
[arXiv:astro-ph/9805201 [astro-ph]].

\bibitem{Riess:2021jrx}
A.~G.~Riess, W.~Yuan, L.~M.~Macri, D.~Scolnic, D.~Brout, S.~Casertano, D.~O.~Jones, Y.~Murakami, L.~Breuval and T.~G.~Brink, \textit{et al.}
Astrophys. J. Lett. \textbf{934}, no.1, L7 (2022)
doi:10.3847/2041-8213/ac5c5b
[arXiv:2112.04510 [astro-ph.CO]].

\bibitem{Ross:2020lqz}
A.~J.~Ross, J.~Bautista, R.~Tojeiro, S.~Alam, S.~Bailey, E.~Burtin, J.~Comparat, K.~S.~Dawson, A.~de Mattia and H.~du Mas des Bourboux, \textit{et al.}
Mon. Not. Roy. Astron. Soc. \textbf{498}, no.2, 2354-2371 (2020)
doi:10.1093/mnras/staa2416
[arXiv:2007.09000 [astro-ph.CO]].


\bibitem{Seikel:2012}
M.~Seikel, S.~Yahya, R.~Maartens, and R.~Clarkson,
Phys. Rev. D \textbf{86}, 083001 (2012)
doi:10.1103/PhysRevD.86.083001
[arXiv:astro-ph/1205.3431 [astro-ph]].

\bibitem{Simon:2004tf}
J.~Simon, L.~Verde and R.~Jimenez,
Phys. Rev. D \textbf{71}, 123001 (2005)
doi:10.1103/PhysRevD.71.123001
[arXiv:astro-ph/0412269 [astro-ph]].

\bibitem{Stern:2009ep}
D.~Stern, R.~Jimenez, L.~Verde, M.~Kamionkowski and S.~A.~Stanford,
JCAP \textbf{02}, 008 (2010)
doi:10.1088/1475-7516/2010/02/008
[arXiv:0907.3149 [astro-ph.CO]].

\bibitem{Thomas:2009wg}
D.~Thomas, C.~Maraston, K.~Schawinski, M.~Sarzi and J.~Silk,
Mon. Not. Roy. Astron. Soc. \textbf{404}, 1775 (2010)
doi:10.1111/j.1365-2966.2010.16427.x
[arXiv:0912.0259 [astro-ph.CO]].

\bibitem{Tomasetti:2023kek}
E.~Tomasetti, M.~Moresco, N.~Borghi, K.~Jiao, A.~Cimatti, L.~Pozzetti, A.~C.~Carnall, R.~J.~McLure and L.~Pentericci,
[arXiv:2305.16387 [astro-ph.CO]].

\bibitem{Vagnozzi:2021tjv}
S.~Vagnozzi, F.~Pacucci and A.~Loeb,
JHEAp \textbf{36}, 27-35 (2022)
doi:10.1016/j.jheap.2022.07.004
[arXiv:2105.10421 [astro-ph.CO]].

\bibitem{Verde:2019ivm}
L.~Verde, T.~Treu and A.~G.~Riess,
Nature Astron. \textbf{3}, 891
doi:10.1038/s41550-019-0902-0
[arXiv:1907.10625 [astro-ph.CO]].

\bibitem{Wang:2018}
L.-L.~Wang, A.-L.~Luo, S.-Y.~Shen, W.~Hou, X.~Kong, Y.-H.~Song, J.-N.~Zhang, H.~Wu, Z.-H.ZCao, Y.-H.~Hou, Y.-F.~Wang, Y.~Zhang, Y.-H.Zhao,
Mon. Not. Roy. Astron. Soc. \textbf{474}, no.2, 1873-1885 (2018)
10.1093/mnras/stx2798
[arXiv:1710.10611 [astro-ph.CO]].

\bibitem{Williams:2008zd}
R.~J.~Williams, R.~F.~Quadri, M.~Franx, P.~van Dokkum and I.~Labbe,
Astrophys. J. \textbf{691}, 1879-1895 (2009)
doi:10.1088/0004-637X/691/2/1879
[arXiv:0806.0625 [astro-ph]].

\bibitem{Worthey:1994iw}
G.~Worthey,
Astrophys. J. Suppl. \textbf{95}, 107-149 (1994)
doi:10.1086/192096

\bibitem{Yu:2017iju}
H.~Yu, B.~Ratra and F.~Y.~Wang,
Astrophys. J. \textbf{856}, no.1, 3 (2018)
doi:10.3847/1538-4357/aab0a2
[arXiv:1711.03437 [astro-ph.CO]].

\bibitem{Zhang:2012mp}
C.~Zhang, H.~Zhang, S.~Yuan, T.~J.~Zhang and Y.~C.~Sun,
Res. Astron. Astrophys. \textbf{14}, no.10, 1221-1233 (2014)
doi:10.1088/1674-4527/14/10/002
[arXiv:1207.4541 [astro-ph.CO]].

\bibitem{Zhao:2017cud}
G.~B.~Zhao, M.~Raveri, L.~Pogosian, Y.~Wang, R.~G.~Crittenden, W.~J.~Handley, W.~J.~Percival, F.~Beutler, J.~Brinkmann and C.~H.~Chuang, \textit{et al.}
Nature Astron. \textbf{1}, no.9, 627-632 (2017)
doi:10.1038/s41550-017-0216-z
[arXiv:1701.08165 [astro-ph.CO]].

\bibitem{Zucca:2009nr}
E.~Zucca, S.~Bardelli, M.~Bolzonella, G.~Zamorani, O.~Ilbert, L.~Pozzetti, M.~Mignoli, K.~Kovac, S.~Lilly and L.~Tresse, \textit{et al.}
Astron. Astrophys. \textbf{508}, 1217 (2009)
doi:10.1051/0004-6361/200912665
[arXiv:0909.4674 [astro-ph.CO]].

\end{thebibliography}
\end{document}